\documentclass[sigconf, nonacm]{acmart}

\usepackage{algorithm}

\usepackage{amsmath, amssymb, amsthm}
\usepackage[subrefformat=parens]{subcaption}
\usepackage{tikz-cd}
\usepackage{multirow} 
\usepackage{amsthm}
\newtheorem{claim}{Claim}
\theoremstyle{remark}

\usepackage{float}
\usepackage{dblfloatfix}
\usepackage{placeins}
\usepackage{cuted}   
\usepackage{capt-of} 
\usepackage{graphicx}

\usepackage{algpseudocode}
\newcommand\vldbdoi{XX.XX/XXX.XX}
\newcommand\vldbpages{XXX-XXX}
\newcommand\vldbvolume{14}
\newcommand\vldbissue{1}
\newcommand\vldbyear{2020}
\newcommand\vldbauthors{\authors}
\newcommand\vldbtitle{\shorttitle} 
\newcommand\vldbavailabilityurl{URL_TO_YOUR_ARTIFACTS}
\newcommand\vldbpagestyle{plain} 

\begin{document}
\title{Scalable Maxflow Processing for Dynamic Graphs}


\author{Shruthi Kannappan}
\affiliation{%
  \institution{Indian Institute of Technology Madras}
  \city{Chennai}
  \country{India}
}
\email{shruthikannappan@gmail.com}

\author{Ashwina Kumar}
\affiliation{%
  \institution{Indian Institute of Technology Madras}
  \city{Chennai}
  \country{India}
}
\email{ashwinakumar825@gmail.com}

\author{Rupesh Nasre}
\affiliation{%
  \institution{Indian Institute of Technology Madras}
  \city{Chennai}
  \country{India}
}
\email{rupesh@cse.iitm.ac.in}

\begin{abstract}
The Maximum Flow (Max-Flow) problem is a cornerstone in graph theory and combinatorial optimization, aiming to determine the largest possible flow from a designated source node to a sink node within a capacitated flow network. It has extensive applications across diverse domains such as computer networking, transportation systems, and image segmentation. The objective is to maximize the total throughput while respecting edge capacity constraints and maintaining flow conservation at all intermediate vertices.

Among the various algorithms proposed for solving the Max-Flow problem, the Push–Relabel algorithm is particularly notable for its efficiency and suitability for parallelization, owing to its localized vertex-based operations. This property has motivated extensive research into GPU-accelerated Max-Flow computation, leveraging the high degree of parallelism inherent to modern GPU architectures.

In this paper, we present a novel GPU-parallel Max-Flow algorithm capable of incrementally recomputing the maximum flow of a dynamic graph following a batch of edge updates. In addition, we introduce a high-performance static GPU algorithm designed for efficiently computing the initial Max-Flow on static graphs. We further describe a series of CUDA-specific implementation optimizations that enhance performance, scalability, and memory efficiency on GPU platforms.
\end{abstract}
\maketitle

\pagestyle{\vldbpagestyle}
\begingroup\small\noindent\raggedright\textbf{PVLDB Reference Format:}\\
\vldbauthors. \vldbtitle. PVLDB, \vldbvolume(\vldbissue): \vldbpages, \vldbyear.\\
\href{https://doi.org/\vldbdoi}{doi:\vldbdoi}
\endgroup
\begingroup
\renewcommand\thefootnote{}\footnote{\noindent
This work is licensed under the Creative Commons BY-NC-ND 4.0 International License. Visit \url{https://creativecommons.org/licenses/by-nc-nd/4.0/} to view a copy of this license. For any use beyond those covered by this license, obtain permission by emailing \href{mailto:info@vldb.org}{info@vldb.org}. Copyright is held by the owner/author(s). Publication rights licensed to the VLDB Endowment. \\
\raggedright Proceedings of the VLDB Endowment, Vol. \vldbvolume, No. \vldbissue\ %
ISSN 2150-8097. \\
\href{https://doi.org/\vldbdoi}{doi:\vldbdoi} \\
}\addtocounter{footnote}{-1}\endgroup

\ifdefempty{\vldbavailabilityurl}{}{
\vspace{.3cm}
\begingroup\small\noindent\raggedright\textbf{PVLDB Artifact Availability:}\\
The source code, data, and/or other artifacts have been made available at \url{https://github.com/ShruthiKannappan/dyn_maxflow}.
\endgroup
}

\section{Introduction} \label{sec intro}
Graphs have emerged as a critical data structure for modeling relationships in a wide range of real-world applications, from social networks and recommendation systems to transportation and telecommunication infrastructures. With the exponential growth of data, these graphs have become massive, often containing millions of vertices and billions of edges. Processing such large-scale graphs using sequential algorithms is computationally expensive and inefficient. As a result, parallelization of graph algorithms has become essential for achieving acceptable performance.

Furthermore, many real-world graphs are \textit{dynamic}---their structure evolves gradually over time due to the addition or deletion of nodes and edges, or updates to edge weights. Efficiently handling such changes without recomputing results from scratch presents significant algorithmic and engineering challenges, especially when it comes to parallel processing.

One such core problem in graph theory is the \textbf{maximum flow problem}, which involves determining the maximum amount of flow that can be sent from a source node to a sink node in a \textit{flow network}, where each edge has a capacity constraint. Applications of maximum flow range from traffic and logistics planning to data routing and power distribution.

Over the decades, numerous sequential and parallel algorithms have been proposed for computing maximum flow, including the well-known Goldberg-Tarjan algorithm \cite{goldberg1988}. With the advent of Graphics Processing Units (GPUs) as high-performance, massively parallel computing platforms, several implementations of maxflow algorithms have been adapted to exploit the capabilities of many-core GPUs. These efforts have led to significant reductions in execution time for static graphs. However, recomputing maximum flow on \textit{dynamic graphs} remains a bottleneck, as most existing GPU-based solutions require a full re-execution upon each change in the graph.

This paper addresses the need for efficient parallel maxflow computation on dynamic graphs. We propose both a \textbf{static} and a \textbf{dynamic} maximum flow algorithm designed to run efficiently on the GPU. Our dynamic algorithm leverages the results of a previously computed maxflow and incrementally updates it in response to batch updates in the graph's edge capacities, thus avoiding costly full recomputations.



Through our experimental evaluation, we demonstrate that the dynamic maxflow algorithm implemented in CUDA significantly outperforms traditional recomputation strategies. Specifically, we observe substantial improvements in both execution time and memory consumption when handling batches of updates to edge capacities. This validates the effectiveness of our incremental update approach and highlights its practical applicability for large-scale, evolving graph datasets.

Our specific contributions are as follows:
\begin{itemize}
    \item We devised dynamic maxflow computaion on GPUs and comapred it with existing state of art work alt-pp~\cite{Khatri2022}.
    \item We complement empirical
    results with a formal proof of correctness for our approach.
    \item  We extend the conventional Compressed Sparse Row (CSR) format into a Bi-Directional CSR (Bi-CSR) representation.
    The Bi-CSR explicitly materializes both outgoing and incoming adjacency lists, ensuring symmetric access to residual capacities.
   
\end{itemize}

The rest of this paper is organized as follows:
Section~\ref{sec back} presents the background of the work.
Section~\ref{sec staticmaxflow} presents the static maxflow algorithm on GPU. Section~\ref{sec dynamic} describes the dynamic maxflow algorithm and how it reuses computations from the static algorithm.
Section~\ref{sec optimization} reports the various implementational optimisations.
Section~\ref{sec experiments} reports experimental evaluations.
Finally, Section~\ref{sec conclusion} concludes the paper and outlines potential directions for future work.

\section{Background} \label{sec back}

\subsection{Maximum Flow Problem}
The \emph{maximum flow problem} is a fundamental question in network flow theory. Given a directed graph $G = (V, E)$ where each edge $(u, v) \in E$ has a non-negative capacity $c(u, v)$, and two distinguished vertices: a \emph{source} $s \in V$ and a \emph{sink} $t \in V$, the objective is to determine the maximum amount of flow that can be sent from $s$ to $t$ such that:

\begin{itemize}
    \item The flow on any edge does not exceed its capacity: $0\leq f(u, v) \leq c(u, v)$,
    \item The flow is conserved at intermediate vertices: for all $v \in V \setminus \{s, t\}$,
    \[
        \sum_{u \in V} f(u, v) = \sum_{w \in V} f(v, w).
    \]
\end{itemize}

The problem has several efficient algorithmic solutions, including the Ford-Fulkerson method, the Edmonds-Karp algorithm, and the Push-Relabel method.

The \textit{maximum-flow problem} is a fundamental combinatorial optimization primitive with far-reaching influence across data management, computer vision, and networked systems. Formally, it seeks the maximum feasible flow between a designated source and sink in a capacitated graph while preserving flow conservation constraints. Owing to its generality, the max-flow abstraction has been extensively leveraged in diverse domains, including:

\begin{itemize}
    \item \textbf{Network Routing:} Modeling bandwidth allocation and congestion control in communication and transportation infrastructures.
    \item \textbf{Image Segmentation:} Enabling graph-cut--based delineation of objects and regions in large-scale visual analytics pipelines.
    \item \textbf{Bipartite Matching:} Powering assignment and scheduling frameworks for job placement, resource allocation, and recommendation systems.
    \item \textbf{Circulation and Logistics:} Optimizing multi-commodity flow, routing, and supply-chain distribution in large-scale planning systems.
\end{itemize}

From a data-systems perspective, max-flow serves as a critical kernel for graph analytics frameworks and parallel query execution engines, motivating efficient GPU-accelerated and incremental variants studied in this work.

\subsection{Related Works}
The maximum flow problem has seen substantial algorithmic progress over 
the decades. The earliest approach, the augmenting path method by Ford and Fulkerson~\cite{ford1956}, finds paths from source to sink along which flow can be pushed until no such paths remain. Though conceptually simple, it is pseudo-polynomial in runtime. The method was refined by Edmonds and Karp~\cite{edmonds1972}, who applied BFS to always choose the shortest augmenting path, leading to a time complexity of $\mathcal{O}(|V||E|^2)$. Dinitz~\cite{dinitz1970} introduced the concept of layered residual graphs and blocking flows, improving the runtime to $\mathcal{O}(|V|^2|E|)$.

A significant conceptual shift occurred with Karzanov~\cite{karzanov1974}, who introduced the idea of preflows—allowing temporary violation of flow conservation—to speed up convergence. Building on this, Goldberg and Tarjan proposed the \textit{Push-Relabel} algorithm~\cite{goldberg1988}, which maintains a preflow and adjusts node heights to direct excess flow toward the sink. This method achieved a worst-case time complexity of $\mathcal{O}(|V|^2|E|)$ and became the standard due to its locality and efficiency.

Push-relabel algorithms are particularly well-suited to parallelization since the push and relabel operations can often be performed independently across nodes. However, most existing parallel implementations~\cite{anderson1995} rely on locking mechanisms to manage concurrency, which is unsuitable for GPU architectures that require lock-free, massively parallel designs such as those supported by CUDA.
 Hong~\cite{hong2008} proposed a lock-free push-relabel algorithm that leverages atomic operations, such as \emph{fetch-and-add}, to enable fine-grained parallelism without the overhead of locks. 

Although the lock-free push-relabel algorithm eliminates locking overhead and enables fine-grained parallelism, it is  slow in practice because threads may perform many unnecessary push and relabel operations, leading to inefficient progress. Previous studies have shown that the practical performance of the push-relabel algorithm can be significantly improved by employing two heuristics: \emph{global relabeling} and \emph{gap relabeling}. The height $h$ of a vertex guides the direction of flow towards the sink. The global relabeling heuristic updates vertex heights to their shortest distance from the sink using a backward breadth-first search (BFS) in the residual graph, while the gap relabeling heuristic identifies gaps in the set of active heights and deactivates unreachable vertices.

Chapter 5 of \emph{GPU Computing Gems Jade Edition}~\cite{wu2012gems} provides a comprehensive overview of efficient CUDA algorithms for the maximum flow problem. It discusses the design and optimization of parallel push-relabel algorithms on GPUs.

Khatri et al. \cite{Khatri2022} (referred to as alt-pp since they perform push and pull in alternate iterations) present several optimizations and approximation techniques for maximum flow computation, including a pull-based algorithm for dynamic scenarios.

\section{GPU-Static-Maxflow Algorithm} \label{sec staticmaxflow}
Drawing inspiration from the improvements by global relabeling, we present a static maxflow Algorithm \ref{static_mf}. Similar to \cite{wu2012gems} push relabel kernel performs pushes and relabels only on active vertices i.e with positive excess ($e(v)>0$) and $height <|V|$.
This algorithm performs the backward BFS with sink as root on GPU to avoid the copies between the GPU and CPU memory. The vertices which cannot reach the sink are assigned a height of $|V|$. At the end of the algorithm, the preflow has no active vertices and the excess at the sink is the maximum flow.

\begin{algorithm}
\caption{GPU-Static-Maxflow}\label{static_mf}
\begin{algorithmic}[1]  
\ForAll{$u\in V$}
\State $e(u) \gets 0$
\State $h(u) \gets 0$ 
\EndFor
\ForAll{$(u, v) \in E$} 
    \State $c_f(u, v) \gets c_{uv}$
    \State $c_f(v, u) \gets c_{vu}$
\EndFor
\ForAll{$(s, u) \in E$}
    \State $c_f(s, u) \gets 0$
    \State $c_f(u, s) \gets c_{us} + c_{su}$
    \State $e(u) \gets c_{su}$
    \State $e(s) \gets e(s) - c_{su}$
\EndFor
\While{$A\neq \phi$} \Comment{$A=\{u\in V-\{s,t\}| e(u)>0 \& h(u)<|V| \}$}
    \State call Backward BFS Algorithm \ref{backward_bfs}
    \State launch a push-relabels kernel \ref{pr_kernel}
    \State launch a remove-invalid-edges kernel \ref{invalid_edges_kernel}
\EndWhile
\State $maxflow\gets e(t)$
\end{algorithmic}
\end{algorithm}

\begin{algorithm}
\caption{push-relabel kernel}\label{pr_kernel}
\begin{algorithmic}[1]  

\State Launch threads for each $u \in V \setminus \{s, t\}$ to execute the following:
\State $cnt\gets 0$
\While{$cnt<KERNEL CYCLES$}
    \If{$h(u)<|V|$ $ \&\& e(u)>0  $}

    \State $e' \gets e(u)$
    \State $\hat{v} \gets \textit{null}$
    \State $\hat{h} \gets \infty$
    \ForAll{$(u, v) \in E_f$ \textbf{where} $c_f(u, v) > 0$}
        \State $h' \gets h(v)$
        \If{$h' <\hat{h}$}
            \State $\hat{v} \gets v$
            \State $\hat{h} \gets h'$
        \EndIf
    \EndFor \Comment{$\hat{v}$ is $u$'s lowest neighbor in $E_f$}
    \If{$h(u) > \hat{h}$} \Comment{push$(u, \hat{v})$ is applicable}
        \State $d \gets \min(e', c_f(u, \hat{v}))$
        \State $c_f(u, \hat{v}) \gets c_f(u, \hat{v}) - d$ \Comment{Atomic fetch and subtract}
        \State $c_f(\hat{v}, u) \gets c_f(\hat{v}, u) + d$ \Comment{Atomic fetch and add}
        \State $e(u) \gets e(u) - d$ \Comment{Atomic fetch and subtract}
        \State $e(\hat{v}) \gets e(\hat{v}) + d$ \Comment{Atomic fetch and add}
    \Else \Comment{lift$(u)$ is applicable}
        \State $h(u) \gets \hat{h} + 1$
    \EndIf
    \EndIf
    \State $cnt++$
    
\EndWhile

\end{algorithmic}
\end{algorithm}

\begin{algorithm}
\caption{remove-invalid-edges kernel}\label{invalid_edges_kernel}
\begin{algorithmic}[1]  

\State Launch threads for each $u \in V \setminus \{s, t\}$ to execute the following:
\For{$(u,v)\in E_f$}
 \If{$h(u)>h(v)+1$}
        \State $e(u) \gets e(u) - c_f(u,v)$ \Comment{Atomic fetch and subtract}
        \State $e(v) \gets e(v) + c_f(u,v)$ \Comment{Atomic fetch and add}
        \State $c_f(v,u) \gets c_f(v,u) + c_f(u,v)$
        \State $c_f(u,v) \gets 0$
\EndIf
\EndFor
\end{algorithmic}
\end{algorithm}

\begin{algorithm}
\caption{Backward BFS}\label{backward_bfs}
\begin{algorithmic}[1]  
\State $h(t) \gets 0$
\ForAll{$u \in V \setminus \{t\}$}
    \State $h(u) \gets |V|$
\EndFor
\State $bfsheight \gets 0$
\State $bfsflag\gets true$
\While{ $bfsflag$}
\State $bfsflag\gets false$
\State Launch kernel for each $u \in V $ to execute the following:
\If{$h(u) == bfsheight$}
    \For {$\ (v,u) \in E_f$}
    \If{$h(v)== |V|$}
        \State $h(v)\gets h(u)+1$
        \State $bfsflag\gets true$
    \EndIf
    \EndFor
\EndIf
\State End kernel
\State $bfsheight++$
\EndWhile
\end{algorithmic}
\end{algorithm}

\subsection{Proof of correctness}

\lemma{At the end of each iteration $h(v)\leq d(v)$, where d(v) represents the shortest distance (in terms of number of edges and not weights) from v to sink(t)}\label{distheight}
\proof After the push relabels kernel is called, the parallel algorithm might violate the height invariant of the serial algorithm i.e $\forall (u,v)\in E_f$, $h(u)\leq h(v)+1$. This invariant could be broken by a scenario such as a vertex v might have read a value of $h(u)$ before it performed a relabel, and hence, the $(u,v)$ edge could be added and the relabel of u could have taken place simultaneously. 
\\ The remove-invalid edges restores this invariant by removing the invalid edges, the reverse edges if added will follow the invariant.
\\ $\forall (u,v)\in E_f$ is same as no steep downward edges.
\\ Let us assume $\exists v$ such that  $d(v)<h(v)$ for some vertex v. This means that there exists a steep downward edge, which is not possible by the invariant. Hence the assumption is false, and we conclude that for every vertex \( v \), the condition \( d(v) \geq h(v) \) holds. 

\theorem{Height of every vertex is non-decreasing }\label{heightinc}
\proof The height of any vertex is changed by the following two operations:
\begin{enumerate}
    \item Relabel - increases the height of the vertex by at least 1.
    \item Backward BFS - never decreases the height, because it assigns h(v) to d(v).  ($\forall u,h(u)\leq d(u)$ from lemma \ref{distheight}) 
\end{enumerate}
\theorem {At the end of the algorithm, we can construct a flow function which is valid on every edge using the residual graph.}
\proof Since the residual edge weights are dealt atomically by the same amount, $\forall u,v\in V$, $c_f(u,v)+c_f(v,u)=c(u,v)+c(v,u)$. Moreover $\forall u,v\in E_f$, $c_f(u,v)\geq 0$, since only pushes by 0 decreases it and the value is always $\leq$ $c_f(u,v)$.
\\ Hence, there are only three possibilities:
\begin{enumerate}
    \item $c_f(u,v)<c(u,v)$ $\implies$ $c_f(v,u)>c(v,u)$, hence we can construct $f(u,v)=c(u,v)-c_f(u,v), f(v,u)=0$.   
    \item $c_f(u,v)>c(u,v)$ $\implies$ $c_f(v,u)<c(v,u)$, hence we can construct $f(v,u)=c(v,u)-c_f(v,u), f(u,v)=0$.
    \item $c_f(u,v)=c(u,v)$ $\implies$ $c_f(v,u)=c(v,u)$, hence we can construct $f(v,u)=0, f(u,v)=0$.
\end{enumerate}
In each of these scenarios $\forall u,v$ $0\leq f(u,v)\leq c(u,v)$,
\theorem{After the termination of the algorithm, a cut A, B can be obtained whose forward edges are saturated by the preflow and backward edges have no flow in them and $C(A,B)$ = $e(t)$}\label{abcut}.
\proof After the algorithm terminates, the vertices of the graph could now be divided into two disjoint sets A, B where $A=\{u|u\in V$  $\&$  $h(v)=|V|\}$ and $B=\{v|v\in V$ 
  $\&$  $h(v)<|V|\}$. 
\\ We can now claim that $\forall (a,b)\in E$ such that $a\in A$ and $b \in B$ $c_f(a,b)=0$, else if $c_f(a,b)>0$ then $h(a)<|V|$ which is a contradiction. Hence $f(a,b)=c(a,b)$
\\Similarly $\forall (a,b)\in E$ such that $a\in A$ and $b \in B$ $f(b,a)=0$, else if $f(b,a)>0$ then $c_f(a,b)>0$ and $h(a)<|V|$ which is a contradiction.
\\Hence the cut A,B is a cut whose forward edges are saturated and backward edges are empty.
Let $e(B)$= $\sum_{u\in B} e(u)$. Since at the termination there are no active vertices, $e(B)$ = $e(t)$.
By definition, 
\\$e(B)$= $\sum_{u\in B} e(u)$ $\Rightarrow$ $\sum_{u\in B} f_{in}(u) - f_{out}(u)$
\\ = $\sum_{u\in B}\sum_{v\in V} f(v,u) - \sum{u\in B}\sum_{v\in V}f(u,v)$
\\ = $\sum_{u\in B}\sum_{v\in A} f(v,u) - \sum{u\in B}\sum_{v\in A}f(u,v)$.
\\From the workout above, $\forall u\in B, v \in A$ $f(u,v) =0$ and $f(v,u) = c(v,u)$.
$\Rightarrow$ $e(B)$ = $C(A,B)$ = $e(t)$.
\lemma{ $\forall v\in V-\{s,t\}$ if $e(v)>0$ then there exists an augmenting path from v to s} \label{path1}
\proof At the termination of the algorithm we can say that there are no vertices such that $e(v)>0$ $\&$ $h(v)<|V|$. Hence if $\exists v\in V-\{s,t\}$ such that $e(v)>0$ then  $h(v)=|V|$ and $v\in A$.
\\Let us assume that $\exists u_0\in A$ such that  $e(u_0)>0$ and there is no path from $u_0$ to s.
Let $X=\{u\in A| u\not\rightsquigarrow s\}$, $Y=\{u\in A| u\rightsquigarrow s\}$.  
\\By assumption, $u_0\in A$. 
Since this is a preflow, $\forall v\in V-\{s,t\}$ $e(v)\geq0$. Let $e(X)=\sum_{u\in X}e(u)$ $=\sum_{u\in X}f_{in}(u)-f_{out}(u)$  
\\$=\sum_{u\in X}\sum_{v\in V}f(v,u)-\sum_{u\in X}\sum_{v\in V}f(u,v)$. 
\\$=\sum_{u\in X}\sum_{v\in X}f(v,u)+ \sum_{u\in X}\sum_{v\in Y}f(v,u)+\sum_{u\in X}\sum_{v\in B}f(v,u)\\-\sum_{u\in X}\sum_{v\in X}f(u,v)-\sum_{u\in X}\sum_{v\in Y}f(u,v)-\sum_{u\in X}\sum_{v\in B}f(u,v)$.
\\$=\sum_{u\in X}\sum_{v\in Y}f(v,u)+\sum_{u\in X}\sum_{v\in B}f(v,u)-\sum_{u\in X}\sum_{v\in Y}f(u,v)\\-\sum_{u\in X}\sum_{v\in B}f(u,v)$.
\\ From Theorem \ref{abcut} , we can see that $\forall v\in B$ and $u\in X ,f(v,u)=0$ $\Rightarrow$ $\sum_{u\in X}\sum_{v\in B}f(v,u) = 0$.
\\Since $u_0\in X$, $e(X)>0$ $\Rightarrow$ $\exists u\in X,v\in Y , f(v,u)>0$ 
\\$\Rightarrow$  $(u,v)\in E_f$,$\Rightarrow$ $u\rightarrow v\rightsquigarrow s$.
Contradiction.
\theorem{After the algorithm terminates then the e(t) is the maxflow}
\proof After the termination of the algorithm, we have a preflow $f$, a cut A, B. 
From $f$, we can construct $f^*$ by sending the excess flow from the overflowing vertices through any augmenting path
to the source which is guaranteed to exist from Lemma \ref{path1}. (Note that these paths lie completely in A; hence, the flow across the  cut (A,B) is not disturbed).
Hence $f^*$ is a valid flow and where $ \forall v\in V-\{s,t\} e(v)=0$,    $-e(s)=e(t)$ = $|f^*|$. 
From Theorem \ref{abcut}, $C(A,B)$ = $e(t)$ = $|f^*|$, by Maxflow-MinCut Theorem \cite{ford1956}, $|f^*|=e(t)$ is the maximum flow of the graph.

\theorem{Algorithm \ref{static_mf} terminates with almost $\mathcal{O}(|V|^2|E|)$ pushes or relabels.}
\proof
From the lemma \ref{heightinc} it follows that each vertex needs at most $|V|$ relabels (relabel during push relabel and relabel during global relabel). Hence, at most, $|V|^2$ relabels could be performed. 
Similar to the proof of the sequential algorithm, there needs to be at least one relabel for a saturated push to repeat on an edge. Hence, for each vertex at most $|V|$ relabels $\implies$ $|V||E|$ saturated pushes
\\ Similarly, the total number of unsaturated pushes  $<4|V|^2(|V|+|E|)$.
\\ The number of pushes and relabels is bounded, hence there will not be any active vertices after these many number of operations,  which implies that the algorithm terminates.
It is to be noted that Global relabel is performed time to time based on $KERNEL\_CYCLES$ parameter, and Global Relabel takes places iff there is some pushes and relabels taking place. Hence Global Relabels is also bounded.
\\\textbf{Note} \begin{enumerate}
    \item Line number 13 in Algorithm \ref{pr_kernel} is different from the lock-free algorithm where $\hat{h}$ is used instead of $h(\hat{v})$ to ensure that the height in one execution is read at most once in order to prove Lemma 3 in ~\cite{hong2008}.
    \item The cut A = $\{u\in V| h(u)=|V|\}$ and B = $\{u\in V| h(u)<|V|\}$ could be used a certificate for the output of maxflow. 
\end{enumerate}

\section{Dynamic Maxflow} \label{sec dynamic}
\subsection{Introduction to dynamic maxflow}
The main focus of this paper is developing dynamic algorithm for maximum flow. 
Dynamic Maxflow refers to recalculating the maxflow through graph G when a batch of edge updates is given. The batch is a set of edges with the new capacities. The new capacities can be of higher or lower capacities than the earlier capacities. 
The aim is to develop an efficient approach that minimizes the computational time while accommodating these changes to the graph and recalculating the maxflow value after these changes. 
\subsection{Dynamic Maxflow Algorithm}
The Dynamic algorithm should be able to continue from the state computed by the static maxflow algorithm.
The Dynamic Maxflow Algorithm \ref{dynamic_mf}, begins with the processing of updates, where the change brought in by the new capacity is added to the residual capacities in parallel.
This could make the residual capacity of some edges negative. This happens when the new capacity is lower than the already existing flow through the edge. To resolve this, flow is sent in reverse direction to make the residual capacity as 0.
Now a recalculation of excess of every vertex is carried out in parallel by launching a thread for each vertex. Again the source, pushes maxflow to saturate all forward edges from itself.
Line 20 to 24 is carried till there are no more $active$ vertices. The definition of $active$ vertex is same where excess is positive and the height $<|V|$.
The difference from static processing is that each deficient vertex is now essentially a sink. Hence all deficient vertices are given height 0 along with the sink and reverse BFS is carried out from this state as shown in Algorithm~\ref{backward_bfs_dyn}. The active vertices carry out the same job but now the flow is directed either to the sink or any of the deficient vertices. Similarly the remove invalid edges is also called. By the end of this procedure. The source along with overflowing vertices will be at height $|V|$ , while the sink and the deficient vertices will be at height 0, the rest vertices could be distributed in various heights. The final maxflow value is computed by simply taking the sum of excess of the vertices at height 0.

\begin{algorithm}
\caption{Dynamic Maxflow}\label{dynamic_mf}
\begin{algorithmic}[1]  
\ForAll{$(u, v) \in Updates$} \Comment{in parallel}
    \State $c_f(u,v) \gets c_f(u,v)+c'_{uv}-c_{uv}$
\EndFor
\ForAll{$u \in V$} \Comment{in parallel}
    \For{$(u,v) \in E$}
    \If{$c_f(u,v)<0$}
        \State $c_f(v,u) \gets c_f(v,u)+c_f(u,v)$
        \State $c_f(u,v) \gets 0$
    \EndIf
    \EndFor
\EndFor
\State calculate excess for each vertex in parallel
\ForAll{$(s, u) \in E$}
    \State $c_f(s, u) \gets 0$
    \State $c_f(u, s) \gets c_{us} + c_{su}$
    \State $e(u) \gets c_{su}$
    \State $e(s) \gets e(s) - c_{su}$
\EndFor
\State $loopflag\gets true$
\While{$loopflag$}
    \State $loopflag\gets false$
    \State call Backward BFS for dynamic maxflow \ref{backward_bfs_dyn}
    \State launch a push-relabels kernel \ref{pr_kernel}
    \State launch a remove-invalid-edges kernel \ref{invalid_edges_kernel}
\EndWhile
\State $maxflow \gets 0$
\ForAll{$u \in V$}
    \If{$h(v)==0$}
        \State $maxflow \gets maxflow + e(v)$
    \EndIf
\EndFor
\end{algorithmic}
\end{algorithm}

\begin{algorithm}
\caption{Backward BFS for dynamic maxflow}\label{backward_bfs_dyn}
\begin{algorithmic}[1]  
\State $h(t) \gets 0$
\State $h(s)\gets |V|$
\For{$u \in V \setminus \{s,t\}$} \Comment{in parallel}
    \If{$e(u)<0$}
        \State $h(u) \gets 0$
    \Else
    \State $h(u) \gets |V|$
    \EndIf
\EndFor
\State $bfsheight \gets 0$
\While{ $bfsheight<|V|$}
\State Launch kernel for each $u \in V $ to execute the following:
\If{$h(u) == bfsheight$}
    \State $\forall (v,u) \in E_f$
    \If{$h(v)== |V|$}
        \State $h(v)\gets h(u)+1$
    \EndIf
\EndIf
\State End kernel
\State $bfsheight++$
\EndWhile

\end{algorithmic}
\end{algorithm}

\subsection{Proof of correctness}
\claim{$\forall v\in V-\{s,t\}$ if $e(v)<0,$ then there exists an augmenting path from t to v}\label{path2}
\proof At the termination of the algorithm, we can say that there are no active vertices(does not include source and sink), i.e. $e(v)>0$ $\&$ $h(v)<|V|$. The negative excess vertices are assigned a height of 0. Hence $\forall v\in V-\{s,t\}$ ($e(v)<0 \rightarrow h(v)=0$ and $v \in B$).
\\Let us assume that $\exists u_0\in B$ such that  $e(u_0)<0$ and there is no path from t to $u_0$ .
\\Let $X'=\{u| u\in B$ and there is no path from t to u$\}$, $Y'=\{u| u\in B$ and there is a path from t to u$\}$. 
\\ From this set constructions we can observe that $X'\cap Y'=\phi$ and $X'\cup Y'=B$,$u_0\in B$. 
\\ Let $e(X')=\sum_{u\in X'}e(u)$ $=\sum_{u\in X'}f_{in}(u)-f_{out}(u)$.
\\$=\sum_{u\in X'}\sum_{v\in V}f(v,u)-\sum_{u\in X'}\sum_{v\in V}f(u,v)$. 
\\$=\sum_{u\in X'}\sum_{v\in X'}f(v,u)+ \sum_{u\in X'}\sum_{v\in Y'}f(v,u)+\sum_{u\in X'}\sum_{v\in A}f(v,u)\\-\sum_{u\in X'}\sum_{v\in X'}f(u,v)-\sum_{u\in X'}\sum_{v\in Y'}f(u,v)-\sum_{u\in X'}\sum_{v\in A}f(u,v)$.
\\$=\sum_{u\in X'}\sum_{v\in Y'}f(v,u)+\sum_{u\in X'}\sum_{v\in A}f(v,u)-\sum_{u\in X'}\sum_{v\in Y'}f(u,v)\\-\sum_{u\in X'}\sum_{v\in A}f(u,v)$.
\\ From Theorem \ref{abcut}, we can see that $\forall v\in A$ and $u\in X' ,f(u,v)=0$, therefore $\sum_{u\in X'}\sum_{v\in A}f(v,u) = 0$.
\\Since $u_0\in X'$, $e(X')<0$ $\Rightarrow$ $\exists u\in X',v\in Y' , f(u,v)>0$ 
\\$\Rightarrow$  $(v,u)\in E_f$,$\Rightarrow$ $t\rightsquigarrow v\rightarrow u$.
Contradiction.
\theorem{After the algorithm terminates then the $maxflow$ is the maximum flow in the graph.}
\proof After the termination of the algorithm, we have a cut A, B. 
From $f$, we can construct $f_1$ by sending the excess flow from overflowing vertices through any augmenting path
to the source which is guaranteed to exist from claim \ref{path1}.(Note that these paths lie completely in A hence the cut (A,B) is not disturbed).
\\ Now $f_1$ is left with vertices with $e(v)<0$, We can now send flow from t to all such v through the augmenting paths which are guaranteed to exist from claim \ref{path2}. On continuing this process we can obtain $f^*$. (Note that these paths lie completely in B hence the cut (A,B) is not disturbed).
\\ Hence $f^*$ is a valid flow and where $C(A,B)$ = $|f^*|$, by maxflow min cut theorem, $f^*$ is a maximum flow for the graph,and A,B is a min cut.
Let $e'$ be the excess of $f^*$, Hence $e'(t)=|f^*|$. The augmentations mean that $e'(t)$ = $\sum_{v\in V-\{s,t\} | e(v)<0 }e(v)+e(t)$ = $maxflow$.
 The algorithm skips this step as e(t) after augmentations is same as e(B), which is the same as the value stored in $maxflow$.
 
\lemma{$h(v)\leq d(v)$, where d(v) represents the shortest distance from v to sink(t) or any deficient vertex}\label{distheight2}
\proof After the push relabels kernel is called, the parallel algorithm might violate the height invariant of the serial algorithm i.e $\forall (u,v)\in E_f$, $h(u)\leq h(v)+1$. This could be possible that v might have read a value of h(u) before it performed a relabel, and hence $(u,v)$ edge could be added and relabel of u could have taken place simultaneously. 
\\ The remove-invalid edges kernel undoes this mistake quickly by removing this invalid edges which will eventually lead to push later if ignored. This just performs the mandatory pushes which will take place before the algorithm terminates and also to ensure the height invariant is maintained. 
\\ If $\forall(u,v)\in E_f, h(u)\leq h(v)+1$, then we do not have any steep downward edges.
\\ For contradiction, let us assume $\exists v$ such that  $d(v)<h(v)$ for some vertex v. This means that there is a steep downward edge which is a contradiction. As the algorithm proceeds some of the deficient vertex might become overflowing or stable ($e(v)=0$), in such a case other deficient vertices and sink will guide the flow. Note that the height does not decrease because of this conversion of deficient to non-deficient since d(v) stores minimum of all such distances to the sink or deficient vertices is chosen, hence d(v) is also increasing.

\theorem{Algorithm \ref{dynamic_mf} terminates.}
\proof The rest of the proof is the same as the static algorithm since it has the same upperbound of the number of pushes, relabels and global relabels.
\\\textbf{Note} The output of the algorithm or execution of the algorithm can be verified with the cut value of A,B where A = $\{u\in V| h(u)=|V|\}$ and B = $\{u\in V| h(u)<|V|\}$

\section{Data Structures and Optimizations} \label{sec optimization}

\subsection{Data Structures}
Efficient neighbor access is critical for GPU-based graph algorithms. We adopt the Compressed Sparse Row (CSR) format for its compactness and coalesced memory access, using two arrays: offset and edgeList. Residual capacities for each edge are maintained in a separate array—referred to as the mirror edgeList—which aligns with the original CSR structure. However, since CSR only encodes outgoing edges, accessing the reverse edge during push-relabel operations is non-trivial. To avoid additional lookups or pointers to reverse edges, we extend the graph representation to a Bi-Directional CSR (Bi-CSR), which explicitly stores both incoming and outgoing edges. For any missing reverse edges, zero-capacity entries are added. This design improves memory locality and eliminates costly indirect accesses, leading to more efficient CUDA execution for flow algorithms. An additional array is maintained which stores the offset of the reverse edges in the CSR as each push operations modifies the forward and reverse edge. This allows us to perform a push on an edge in constant number of memory accesses.
\subsection{Optimisations}


\subsubsection{O1: Graph Processing}
\begin{figure}
    \centering
    \includegraphics[width=\linewidth]{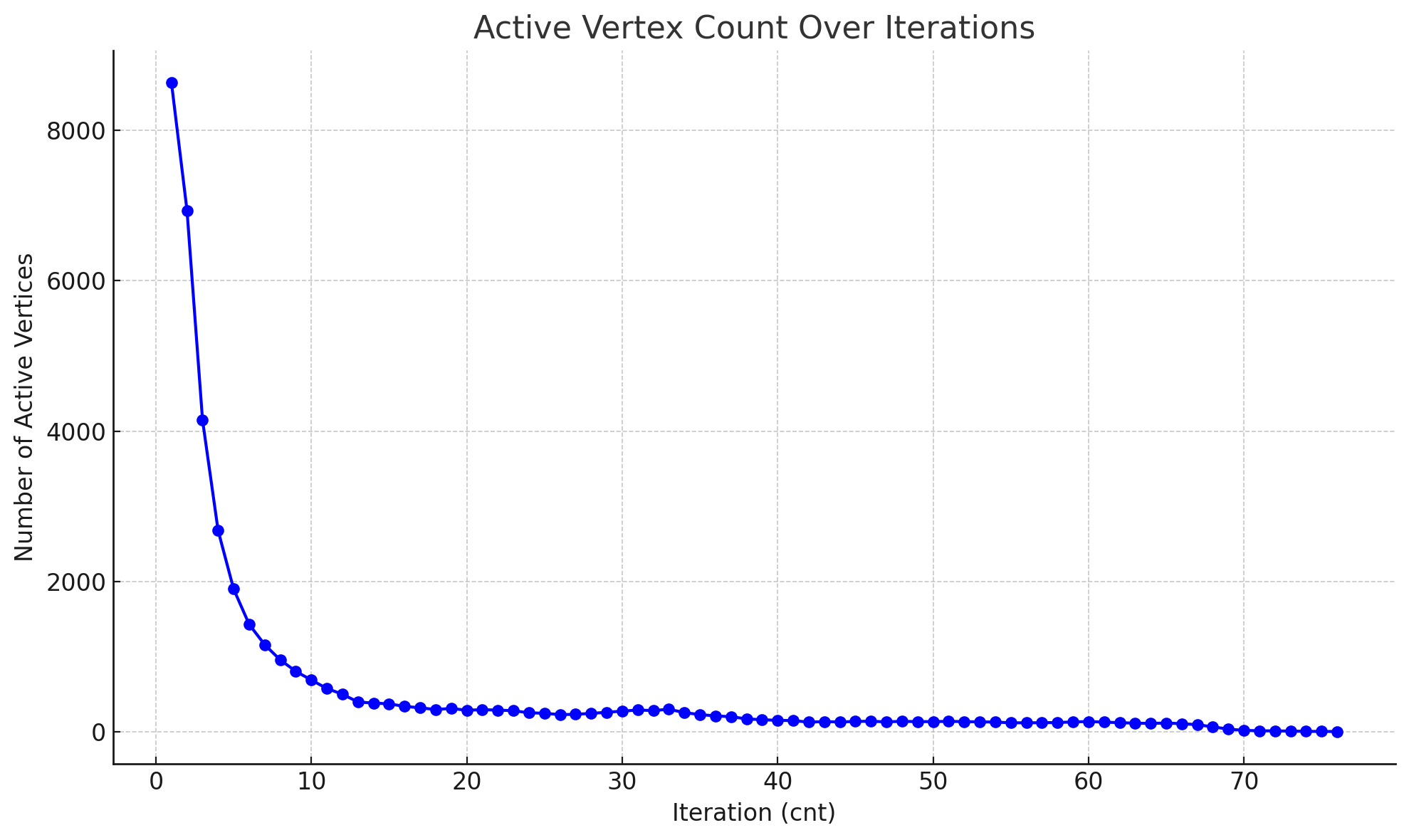}
    \caption{Active vertices in pokec dataset}
    \label{fig:active-vertex}
\end{figure}

The approach of launching a thread for all vertices is called topology-driven approach.
Through experimentation, it is found that the number of active vertices is very low compared to the total number of vertices in the Figure~\ref{fig:active-vertex}.
However in the implementation launching a thread for every vertex might cause unnecessary idle threads occupying the GPU. 
To mitigate this we come up with the data-driven approach of creating a worklist of only active vertices and launching the push-relabel kernel and remove invalid edges kernel only for those vertices which require some work to be done. Hence a worklist of only active vertices is created before the push relabels kernel and only those in the worklist get a thread allocated and processed further. Hence only active vertices number of threads are launched. This optimisation interestingly shows better improvement for the Dynamic Processing  when compared to the Static Implementation. The comparison is shown in the experiments.

\subsubsection{O2: Parallel Push and Pull}
The decremental updates, give rise to deficient vertices i.e $e(u)<0$, which indicates that similar to processing overflowing vertices by pushing flow to neighbours, one can pull flow from neighbors to satisfy the deficiency by increasing the flow in incoming edges. However this requires a different height function, where now steep upward edges are not allowed and the source is at the bottom, and sink at the top, similar to the dynamic maxflow algorithm the overflowing vertices are treated like the source. The theoretical proof of the pull-relabel algorithm can be worked out from the proofs of Static and Maxflow algorithm.
In fact this method has been suggested by \cite{Khatri2022} where push and pull is performed every alternate iteration. However a challenge that remained was proof of termination. Similarly \cite{juntong} creates a different height function and performs push and pull simultaneously. 
However performing push and push pull parallel in our setting is not straightforward. Adopting push and pull in the parallel setting creates the following problems-
\begin{itemize}
    \item The validity of the height functions is required for the proof of termination. It could be possible that an edge is valid in one height function and invalid in the other, while the reverse edge is valid in the opposite. In such a case at least one of the height functions need to be compromised. 
    \item if a deficient and overflowing vertex are neighbors then they might attempt to perform push and pull at the same time, possibly leading to negative residual capacity as both reduce the same quantity. Fixing this would require either locks or some tie breaker.
    \item The positive and negative excess could just end up chasing the other in a cyclical fashion.
\end{itemize}
One could saturate the incoming edges to the sink and treat the resulting deficient vertices as sink in static graph calculation. We have used this approach along with work list creation as static push pull algorithm. The result is the maxflow along with the min-cut. The mincut (S,T) would have the deficient vertices , sink and those vertices which can reach the above former as the T partition. Similarly overflowing vertices, source and vertices which can be reached from the former as the S partition.
When such an algorithm is updated with a batch of changed edges, the maximum flow and min-cut might change, but not significantly compared to the original flow value. 
Using this, we saturate the edges which were across the original min-cut. This saturation relies on the observation that the min-cut might not be too different. The earlier min-cut S,T which had overflowing and deficient vertices respectively and exclusively, now would have both overflowing and deficient vertices on both partitions. Since edges from S to T are saturated, there would be no path from a vertex in S to a vertex in T. Hence keeping overflowing vertices in S at height 0 for pull algorithm and deficient vertices in T at height 0 for push algorithm would ensure that the two algorithms run on disjoint sets of vertices and edges, since the heights are assigned based on BFS from the bases. The two algorithms are run in parallel by two separate workflists and the push and remove push kernels are run in two separate streams to improve parallelization. 
Post the push and pull streams are completed, we are left with a cut on both S and T. 
From the earlier saturated edges and newly saturated edges, we can say that the only possibility that there might be a path from an overflowing vertex to a deficient vertex only from deficient vertex in S and overflowing vertex in T. Hence we launch a push algorithm only on this small section of the graph to save on BFS and other costs.

\section{Experiments}\label{sec experiments}

The following experiments were carried out on NVIDIA Tesla V100-PCIE with the following specifications:
\begin{itemize}
    \item Architecture: Volta
    \item Number of CUDA Cores: 5120
    \item Streaming Multiprocessors (SMs): 80
    \item Total Memory: 32GB
    \item Memory Bandwidth: 900 GB/s.
    
\end{itemize}

\begin{table}[h]
\caption{Input graphs (Original Static Graphs)}
\centering
\small
\setlength{\tabcolsep}{4pt}
\begin{tabular}{l|c|r|r|r|r|r}
\hline
\textbf{Dataset} & \textbf{Short} & \multicolumn{1}{c|}{\textbf{$|$V$|$}} & \multicolumn{1}{c|}{\textbf{$|$E$|$}} & \textbf{Source} & \textbf{Sink} & \textbf{Flow} \\
              & \textbf{name} & \multicolumn{1}{c|}{($\times 10^6$)} & \multicolumn{1}{c|}{($\times 10^6$)} & & & \\
\hline
Pokecwt & PK & 1.6 & 30.6 & 5866 & 5934 & 437500\\ \hline
Flickr-Links   & FR & 1.7 & 15.6 & 397   & 1319705 & 521236  \\ \hline
Stack-overflow & ST & 2.6 & 36.2 & 17034 & 22656   & 1117553 \\ \hline
Wikiwt         & WK & 3.4 & 93.4 & 18646 & 14873   & 250366  \\ \hline
LivJournalwt   & LJ & 4.8 & 69.0 & 10009 & 10029   & 680533  \\ \hline
\end{tabular}
\label{graph-inputs}
\end{table}

The edge weights of above graph are between 1 and 100.

To rigorously evaluate the performance and scalability of the proposed GPU-Static-Maxflow and Dynamic-Maxflow algorithms, we employed a diverse collection of large real-world graph datasets widely used in graph analytics and network-flow benchmarks. Table~\ref{graph-inputs} summarizes the key characteristics of these datasets, including the number of vertices and edges (in millions), the designated source–sink pairs, and the corresponding maximum flow values computed in the static configuration. These graphs—ranging from social networks (e.g., Pokec, LiveJournal) to collaborative and hyperlink networks (e.g., Stack Overflow, Wikipedia, Flickr)—exhibit distinct structural densities and connectivity patterns. Such heterogeneity allows for a comprehensive evaluation of algorithmic behavior under varying graph topologies, degree distributions, and flow magnitudes, ensuring that both the static and dynamic implementations are tested across representative real-world scenarios.
 
\subsection{Experiments on GPU-Static-Maxflow}
In our algorithm, we can indirectly control the number of times global relabel is called by tuning the kernel cycles parameter, which is the maximum number of pushes/relabels a kernel executes for an active vertex. A very low value for this parameter may lead to excessive global relabel calls, resulting in increased computational overhead and decreased performance. Conversely, a high value could lead to flow being directed in the wrong directions. 
We thus thought that the average degree($|E|/|V|$) of the graph is an easy to compute and possible heuristic for kernel cycles that could work in most graphs.
Global relabel can essentially perform the relabeling of many vertices at the same time. Hence if the pushes are completed in the push relabel kernel, we can save up the relabels for the global relabel.

\subsection{Experiments on Dynamic Maxflow}
 We used the heuristic kernel cycles parameter of each graph for the dynamic maxflow code. The update files were generated by randomly choosing the existing edges and assigning a higher or lower weight as per the requirement. In order to observe considerable changes in flow, the edges out of the source and into the sink were chosen with a higher probability than the other edges.
 The dynamic implementations are compared with the alternating push pull improbabilityplementation mentioned in \cite{Khatri2022} as alt-pp. The dynamic implementation is also compared with the time consumed by static algorithm after the underlying graph is updated with the batch of egdes.
\\ Note: $x\%$ updates means the number of updates in a batch is $x*|E|/100$.

\subsubsection{Incremental Updates}
The plots in Figure \ref{fig:inc} shows the performance of the dynamic algorithm when all the updates in the batch is larger than the original capacity.

The dynamic implementations are doing significantly better than the static and alternate approaches especially for incremental. This highlights that dynamic implementations are effective for incremental updates. 
Across all the datasets, O2: push pull streams optimisation seems to be the winner except the pokec dataset, where the topology-driven and data-driven implementations are giving better performance. The topological approach and data-driven approach are giving similar results for incremental. 

\FloatBarrier
\begin{strip}
\centering

\begin{minipage}{0.32\textwidth}\centering
\includegraphics[width=\linewidth,height=0.18\textheight,keepaspectratio]{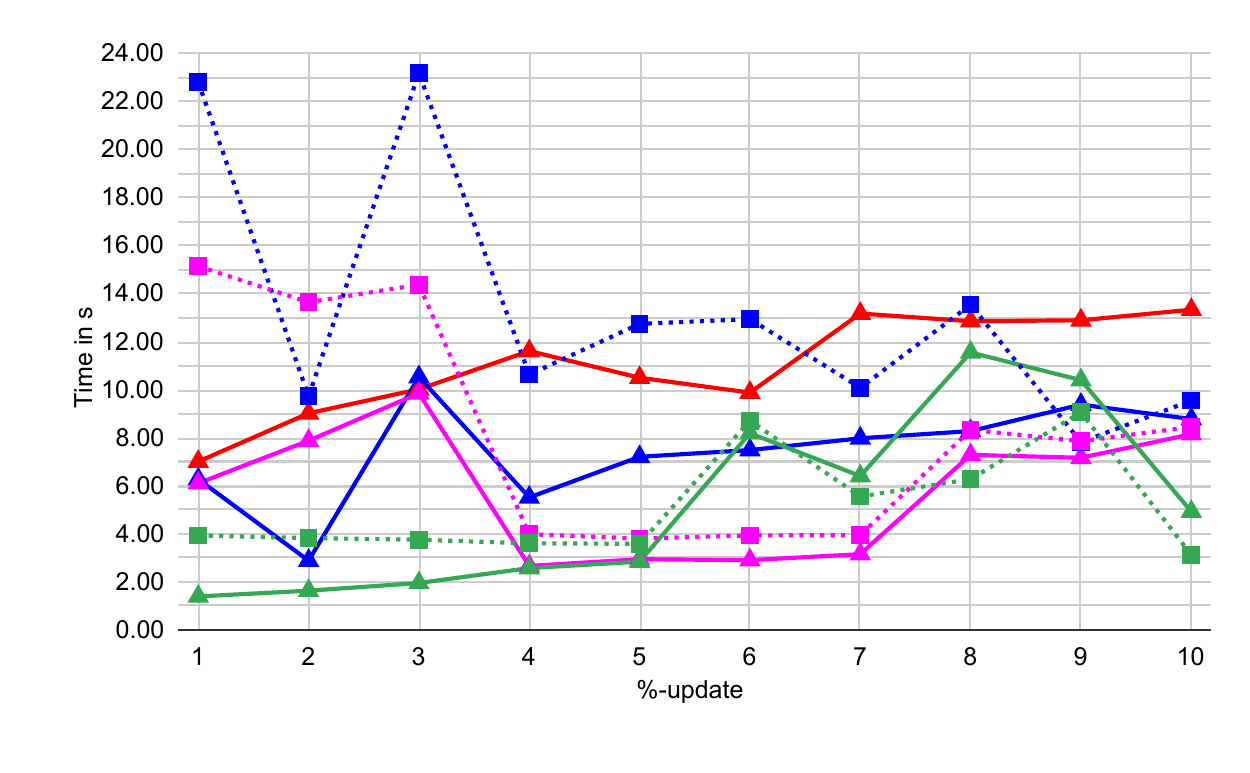}\\[-0.25ex]
{\small (a) Flickr Dataset}
\end{minipage}\hfill
\begin{minipage}{0.32\textwidth}\centering
\includegraphics[width=\linewidth,height=0.18\textheight,keepaspectratio]{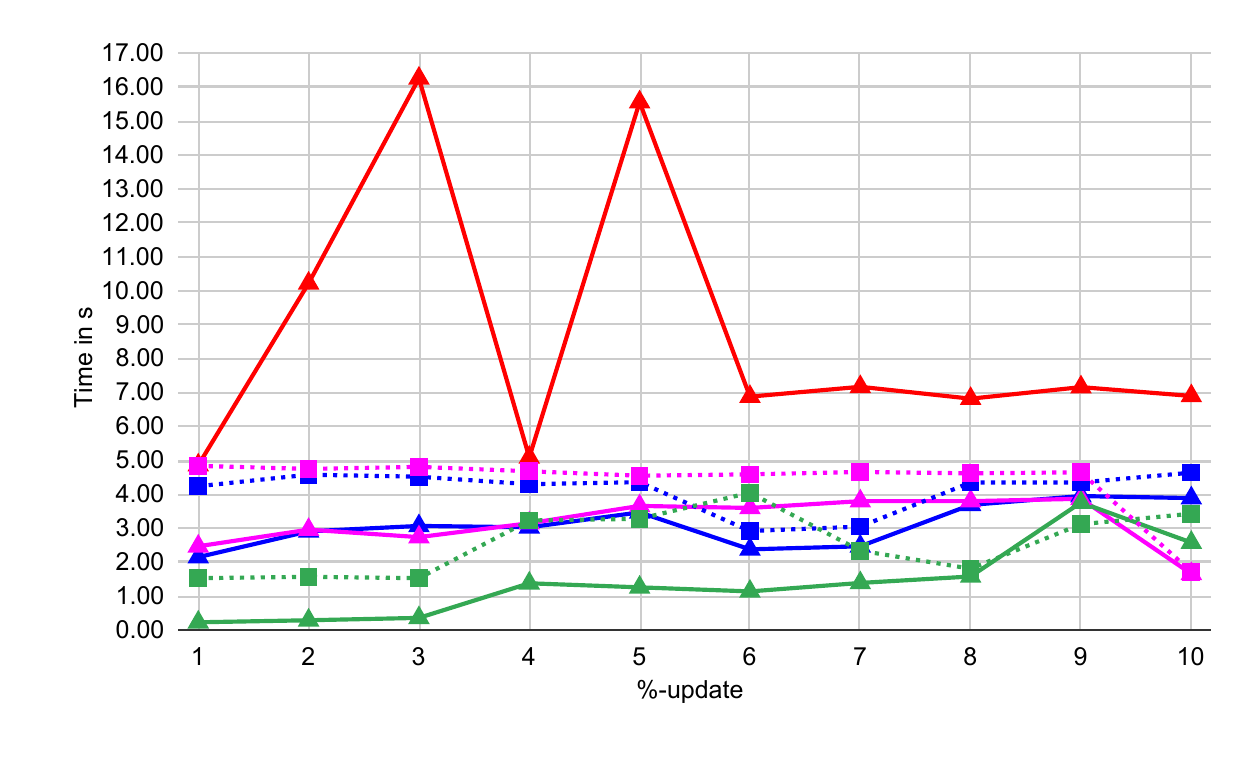}\\[-0.25ex]
{\small (b) LivJournal Dataset}
\end{minipage}\hfill
\begin{minipage}{0.32\textwidth}\centering
\includegraphics[width=\linewidth,height=0.18\textheight,keepaspectratio]{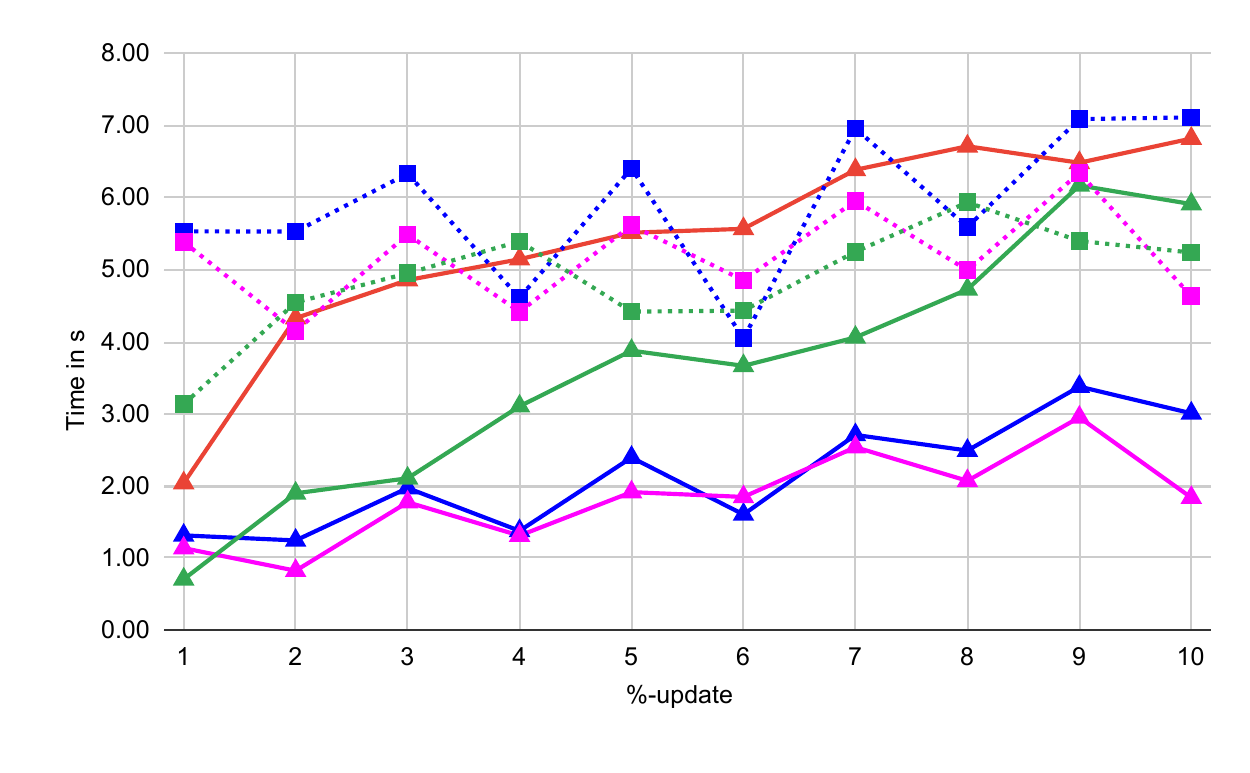}\\[-0.25ex]
{\small (c) Pokecwt Dataset}
\end{minipage}

\vspace{0.5ex}

\begin{minipage}{0.32\textwidth}\centering
\includegraphics[width=\linewidth,height=0.18\textheight,keepaspectratio]{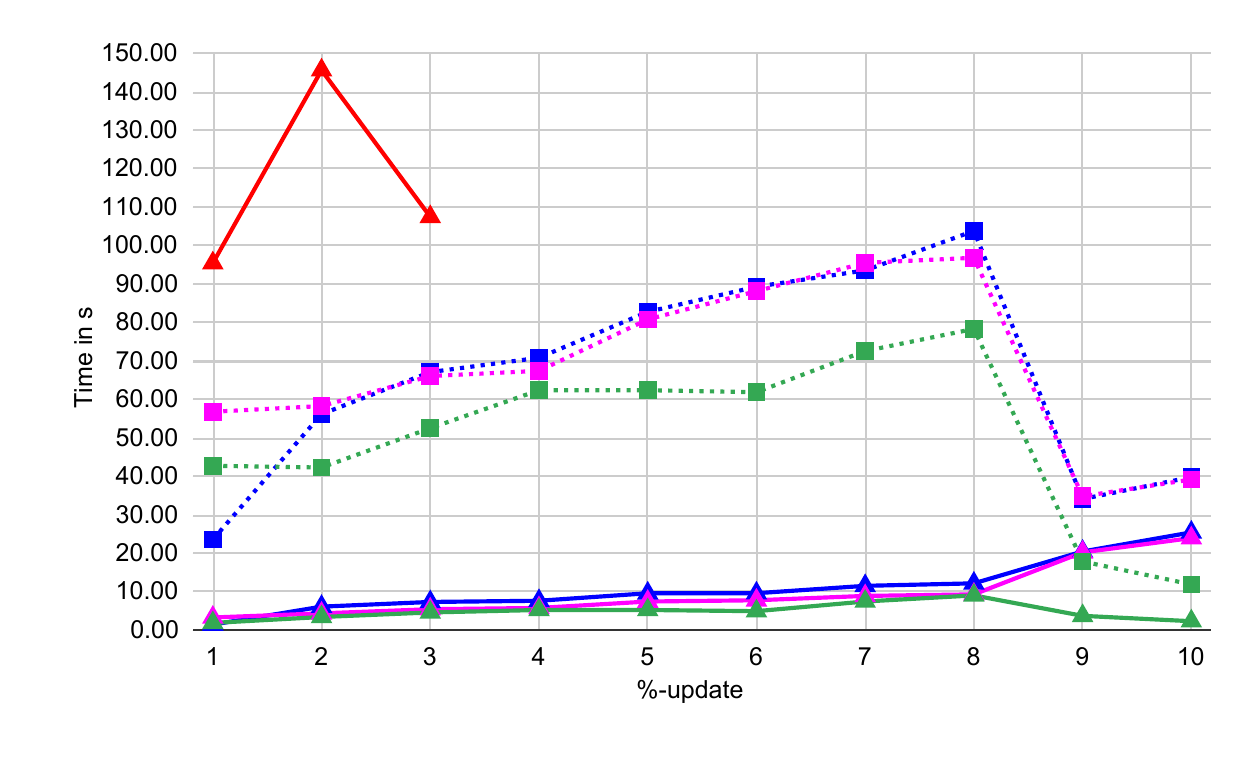}\\[-0.25ex]
{\small (d) Wikipedia Dataset}
\end{minipage}\hfill
\begin{minipage}{0.32\textwidth}\centering
\includegraphics[width=\linewidth,height=0.18\textheight,keepaspectratio]{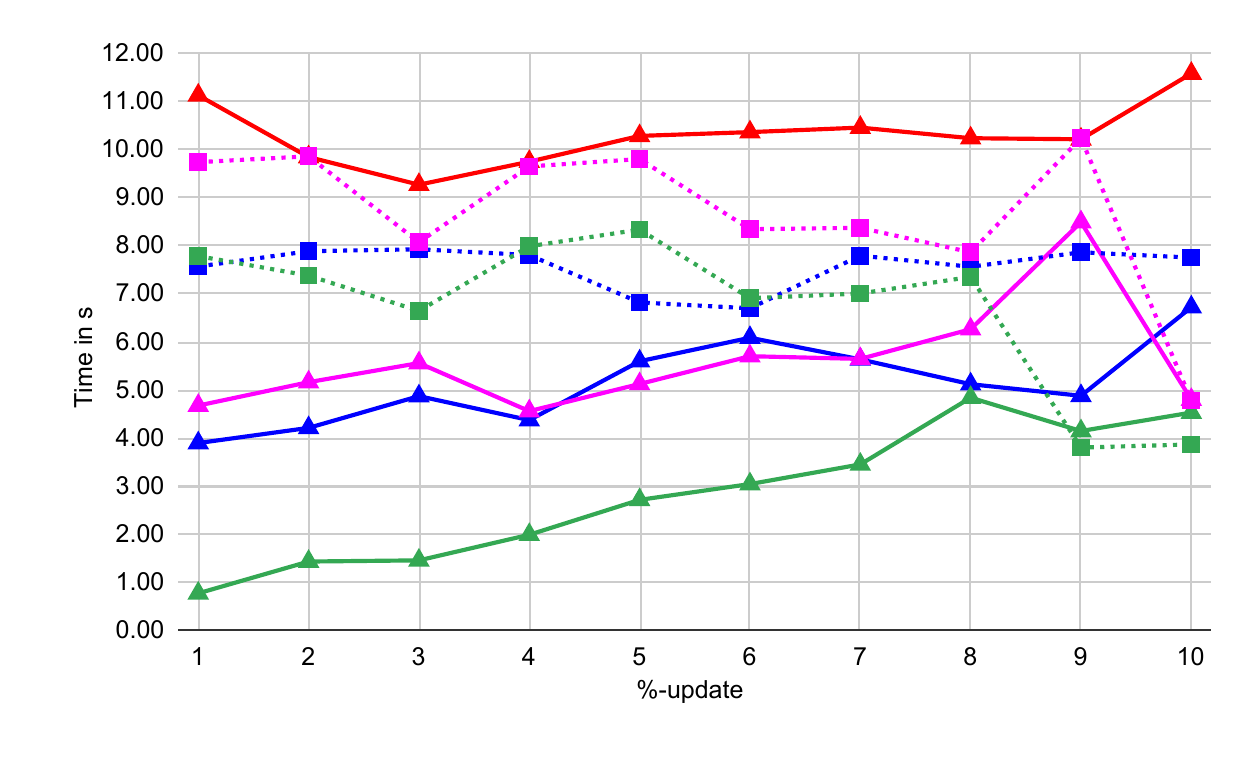}\\[-0.25ex]
{\small (e) Stack Overflow Dataset}
\end{minipage}\hfill
\begin{minipage}{0.15\textwidth}\centering
\includegraphics[width=\linewidth,keepaspectratio]{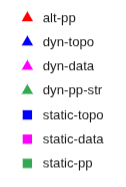}\\[-0.25ex]
{\small Legend}
\end{minipage}

\vspace{0.5ex}
\captionof{figure}{Performance for incremental updates across all datasets. \textit{Legend:} 
\textit{alt-pp} — state-of-the-art dynamic algorithm baseline; 
\textit{dyn-topo} — Dynamic \textit{Push–Relabel} algorithm with topology-based processing; 
\textit{dyn-data} — Dynamic \textit{Push–Relabel} algorithm with data-driven processing; 
\textit{dyn-pp-str} — Dynamic \textit{Push–Pull} algorithm with stream-based data processing; 
\textit{static-topo} — Static \textit{Push–Relabel} algorithm with topology-based processing; 
\textit{static-data} — Static \textit{Push–Relabel} algorithm with data-driven processing; 
\textit{static-pp} — Static \textit{Push–Pull} algorithm with data-driven processing.}
\label{fig:inc}
\end{strip}
\FloatBarrier

\FloatBarrier
\begin{strip}
\centering

\begin{minipage}{0.32\textwidth}\centering
\includegraphics[width=\linewidth,height=0.18\textheight,keepaspectratio]{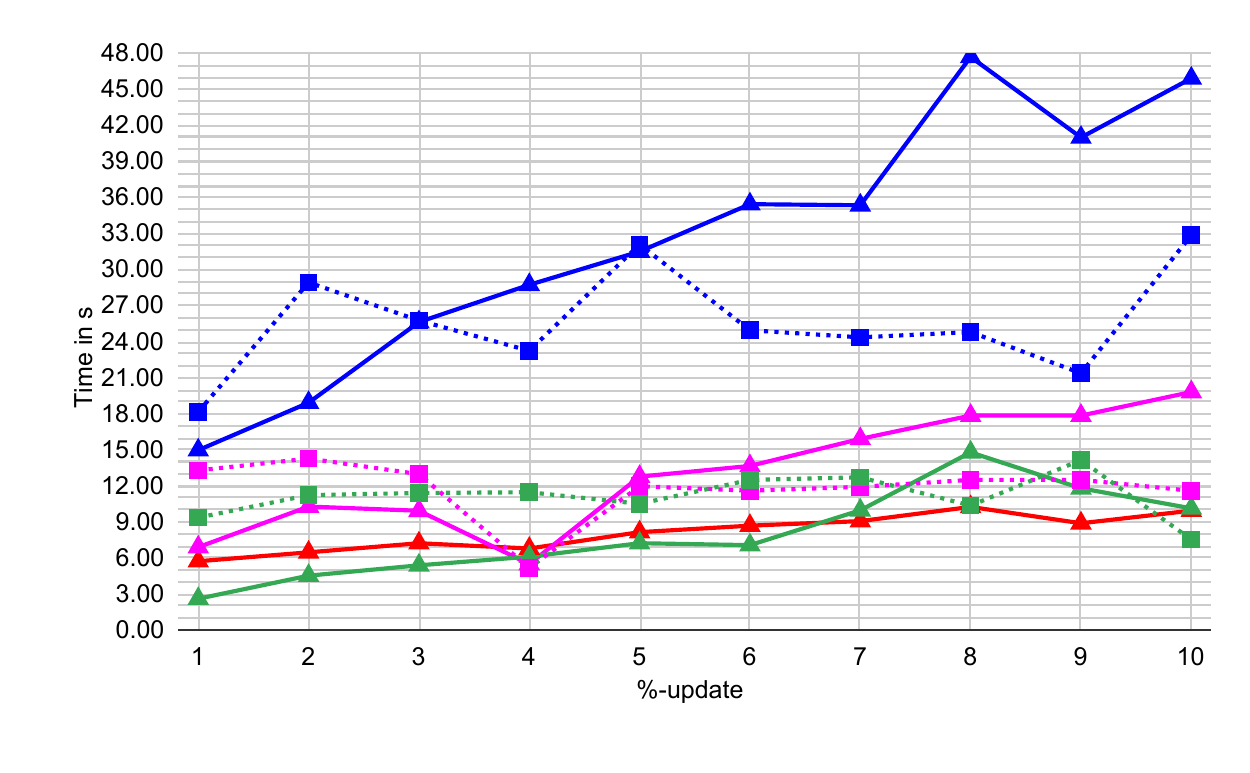}\\[-0.25ex]
{\small (a) Flickr Dataset}
\end{minipage}\hfill
\begin{minipage}{0.32\textwidth}\centering
\includegraphics[width=\linewidth,height=0.18\textheight,keepaspectratio]{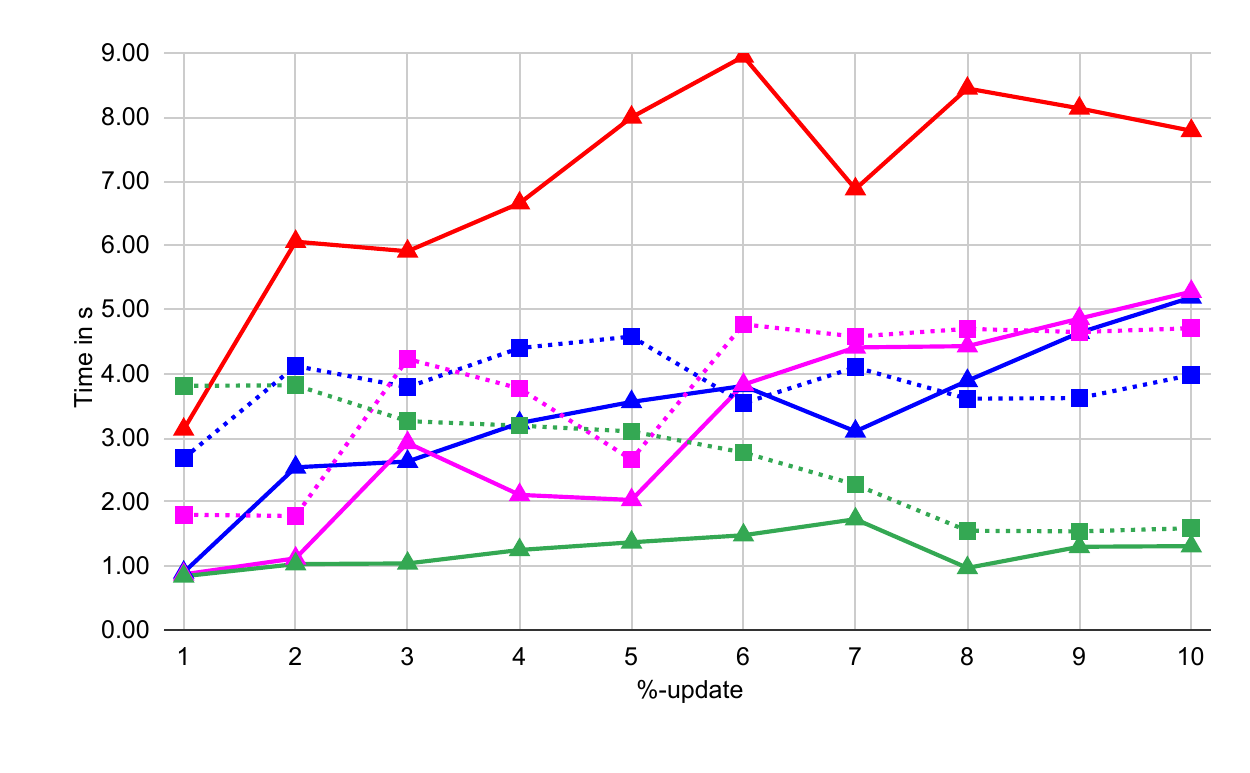}\\[-0.25ex]
{\small (b) LivJournal Dataset}
\end{minipage}\hfill
\begin{minipage}{0.32\textwidth}\centering
\includegraphics[width=\linewidth,height=0.18\textheight,keepaspectratio]{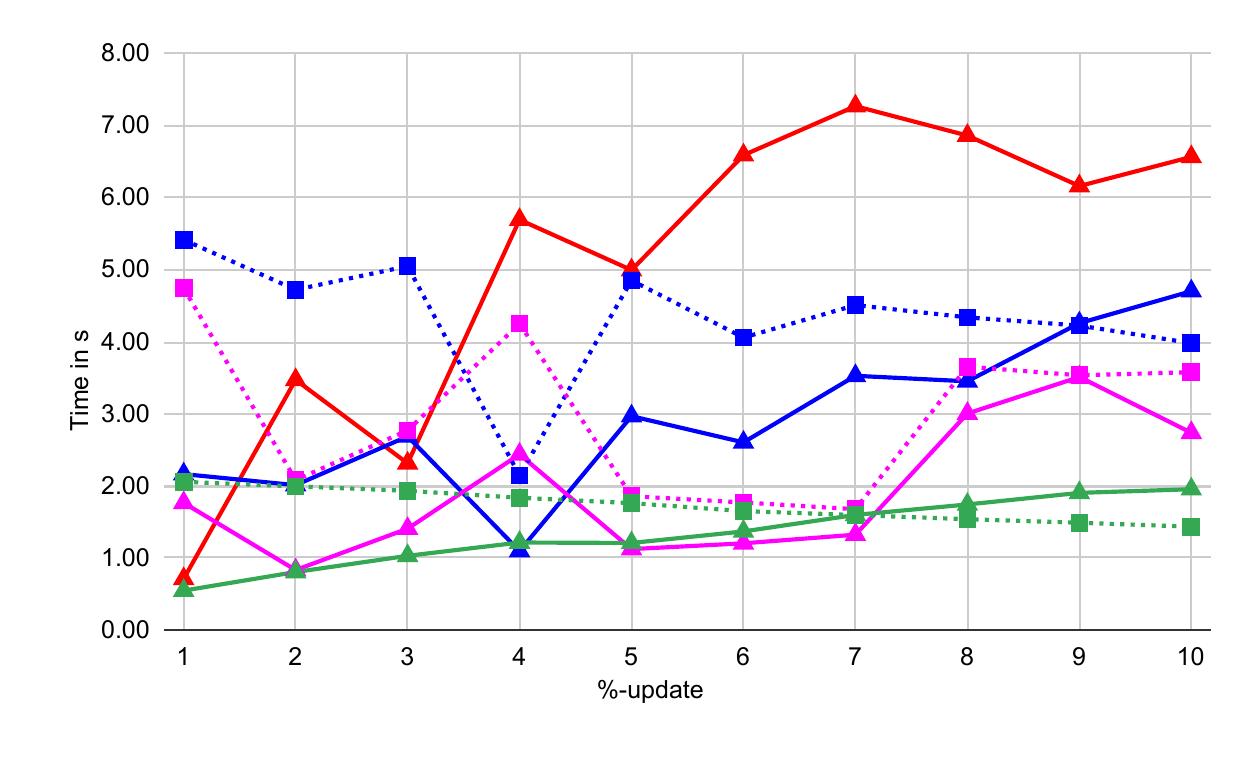}\\[-0.25ex]
{\small (c) Pokecwt Dataset}
\end{minipage}

\vspace{0.5ex}

\begin{minipage}{0.32\textwidth}\centering
\includegraphics[width=\linewidth,height=0.18\textheight,keepaspectratio]{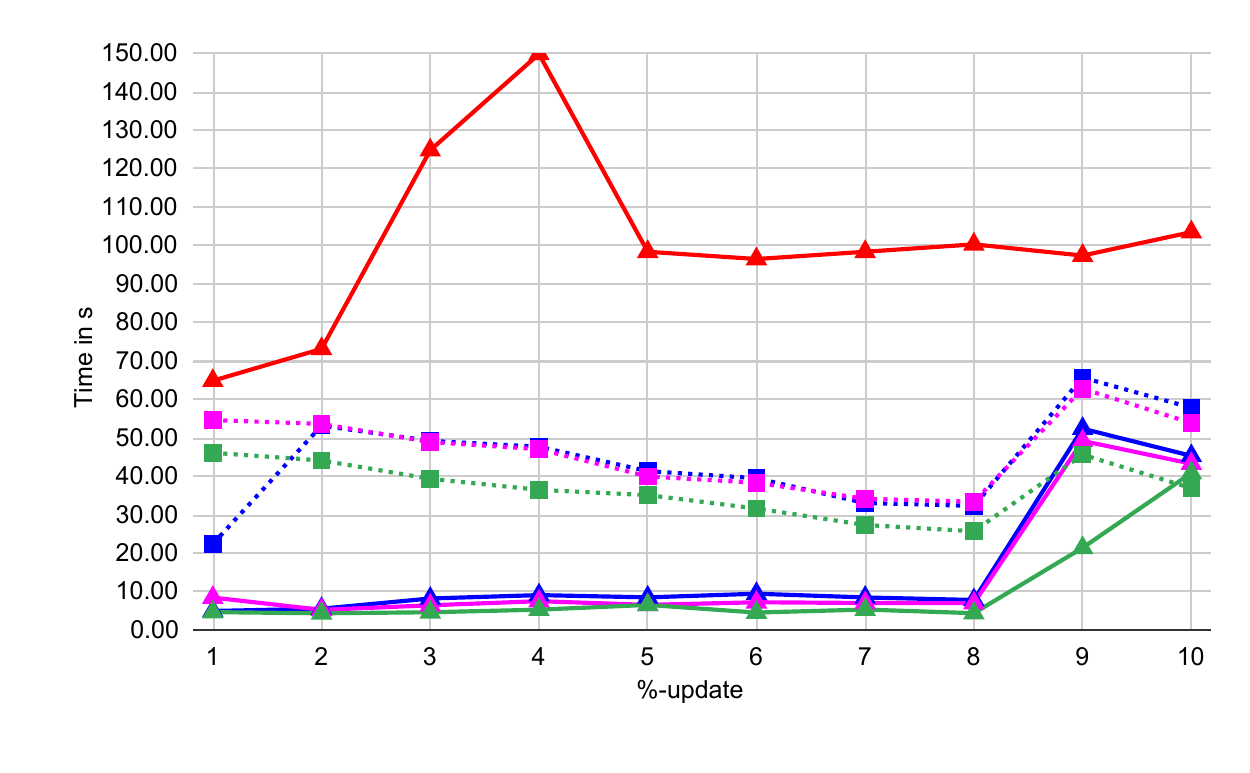}\\[-0.25ex]
{\small (d) Wikipedia Dataset}
\end{minipage}\hfill
\begin{minipage}{0.32\textwidth}\centering
\includegraphics[width=\linewidth,height=0.18\textheight,keepaspectratio]{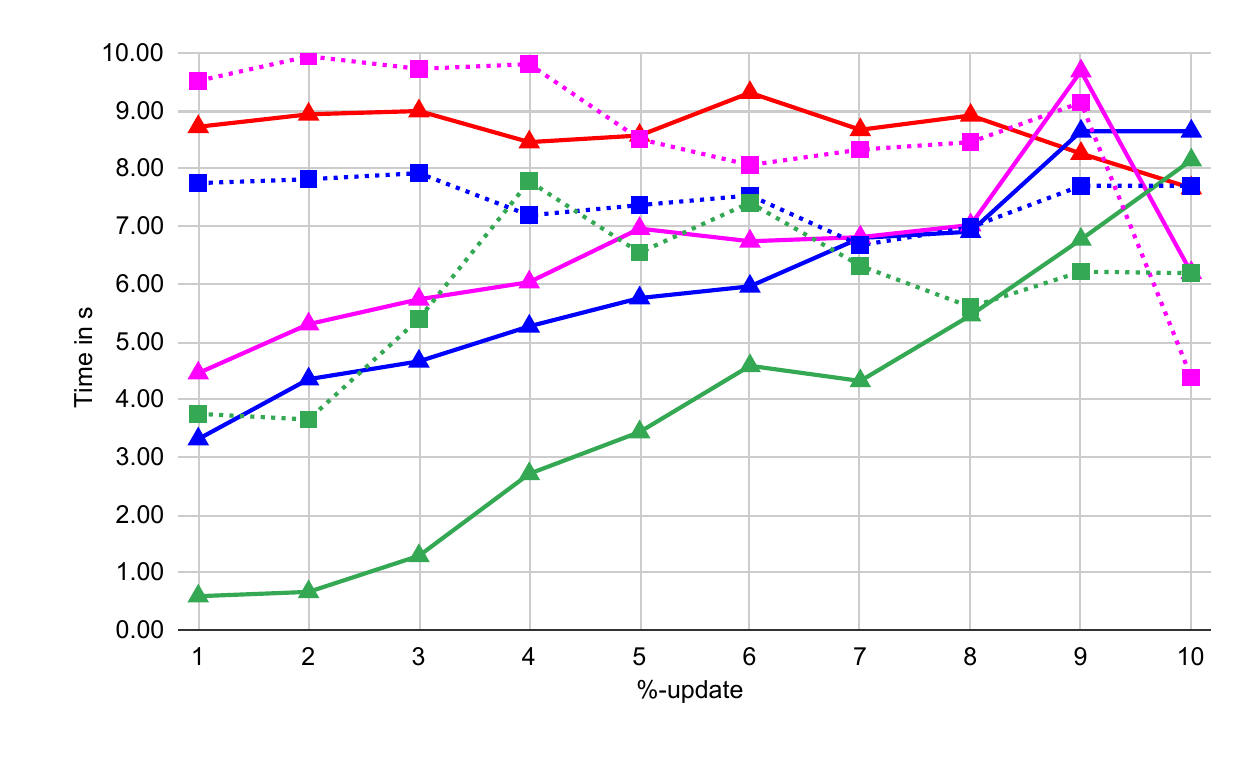}\\[-0.25ex]
{\small (e) Stack Overflow Dataset}
\end{minipage}\hfill
\begin{minipage}{0.15\textwidth}\centering
\includegraphics[width=\linewidth,keepaspectratio]{plots-paper/legend.png}\\[-0.25ex]
{\small Legend}
\end{minipage}

\vspace{0.5ex}
\captionof{figure}{Performance for decremental updates across all datasets.
\textit{Legend:} 
\textit{alt-pp} — state-of-the-art dynamic algorithm baseline; 
\textit{dyn-topo} — Dynamic \textit{Push–Relabel} algorithm with topology-based processing; 
\textit{dyn-data} — Dynamic \textit{Push–Relabel} algorithm with data-driven processing; 
\textit{dyn-pp-str} — Dynamic \textit{Push–Pull} algorithm with stream-based data processing; 
\textit{static-topo} — Static \textit{Push–Relabel} algorithm with topology-based processing; 
\textit{static-data} — Static \textit{Push–Relabel} algorithm with data-driven processing; 
\textit{static-pp} — Static \textit{Push–Pull} algorithm with data-driven processing.}
\label{fig:dec}
\end{strip}
\FloatBarrier

\FloatBarrier
\begin{strip}
\centering

\begin{minipage}{0.32\textwidth}\centering
\includegraphics[width=\linewidth,height=0.18\textheight,keepaspectratio]{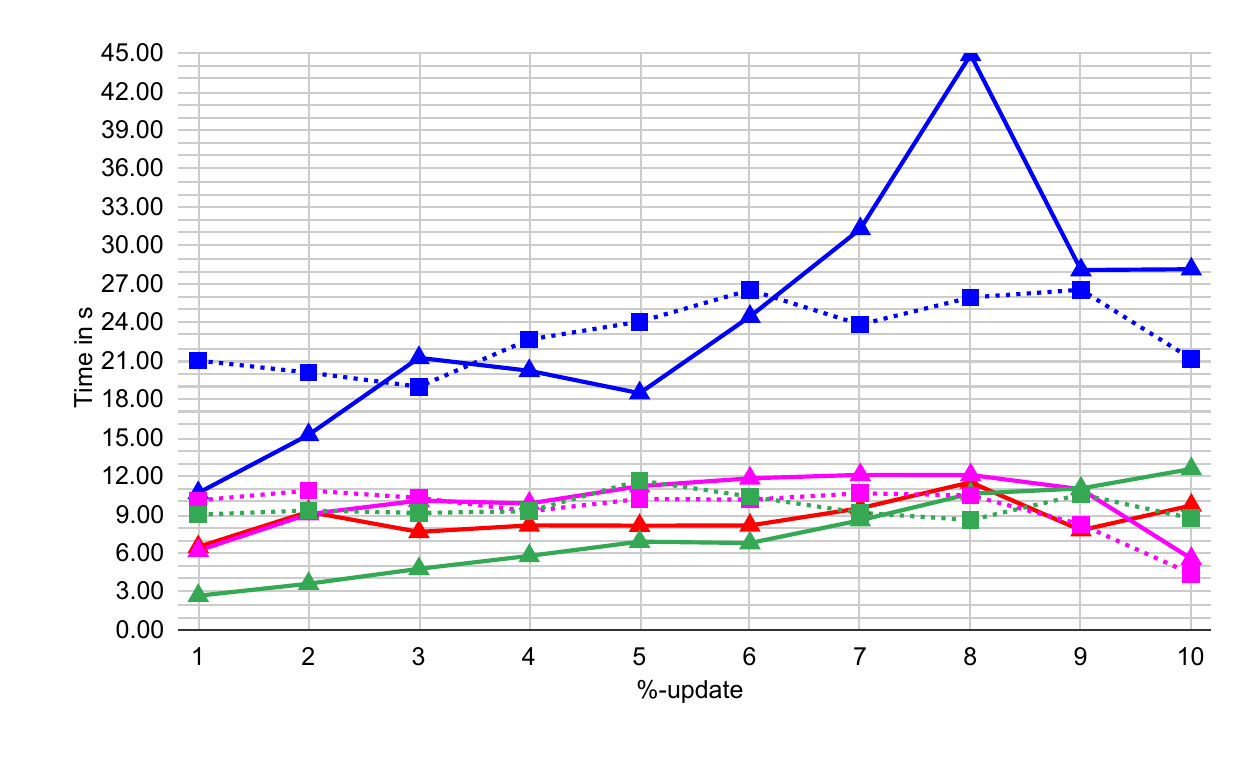}\\[-0.25ex]
{\small (a) Flickr Dataset}
\end{minipage}\hfill
\begin{minipage}{0.32\textwidth}\centering
\includegraphics[width=\linewidth,height=0.18\textheight,keepaspectratio]{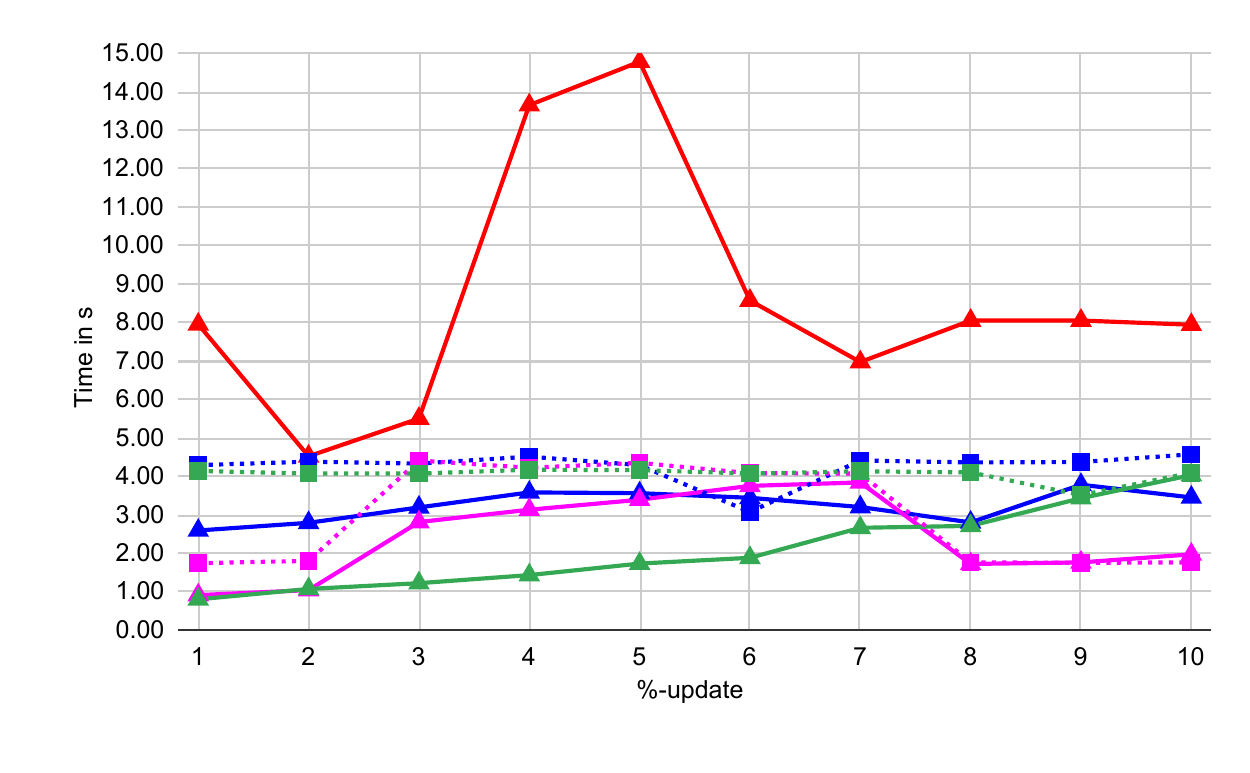}\\[-0.25ex]
{\small (b) LivJournal Dataset}
\end{minipage}\hfill
\begin{minipage}{0.32\textwidth}\centering
\includegraphics[width=\linewidth,height=0.18\textheight,keepaspectratio]{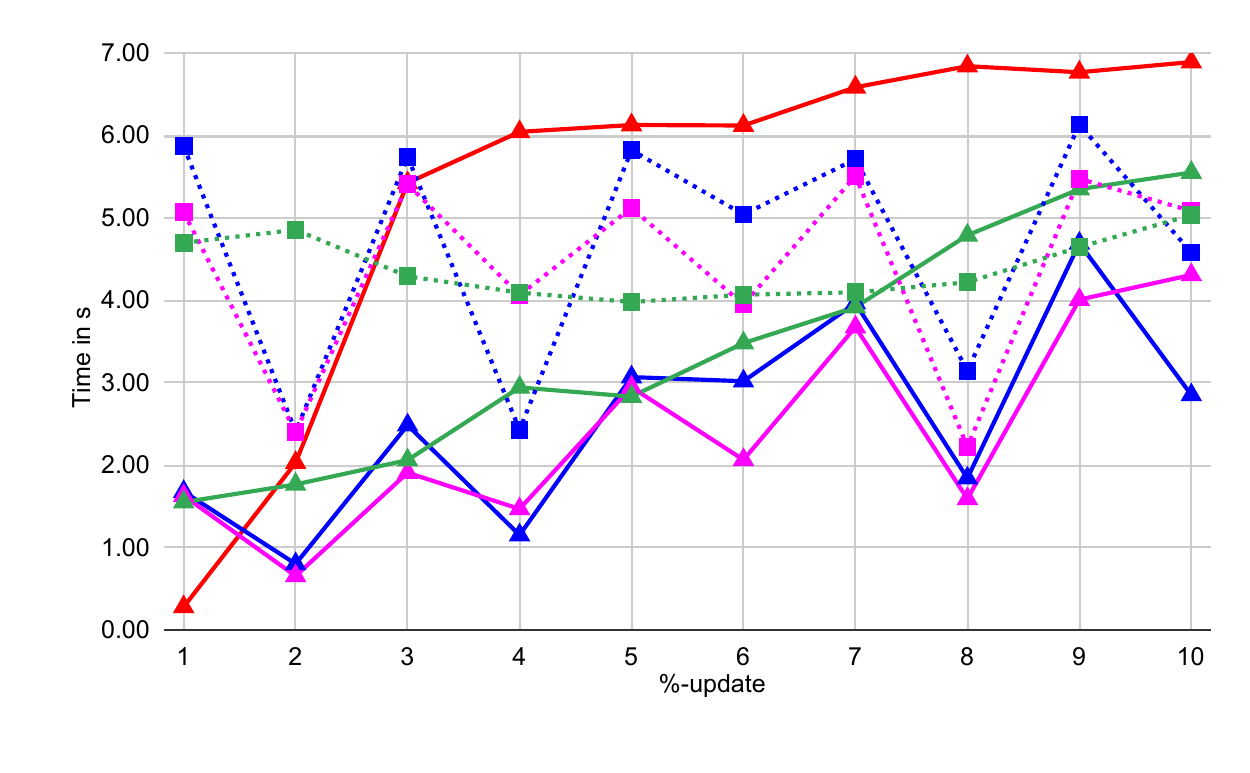}\\[-0.25ex]
{\small (c) Pokecwt Dataset}
\end{minipage}

\vspace{0.5ex}

\begin{minipage}{0.32\textwidth}\centering
\includegraphics[width=\linewidth,height=0.18\textheight,keepaspectratio]{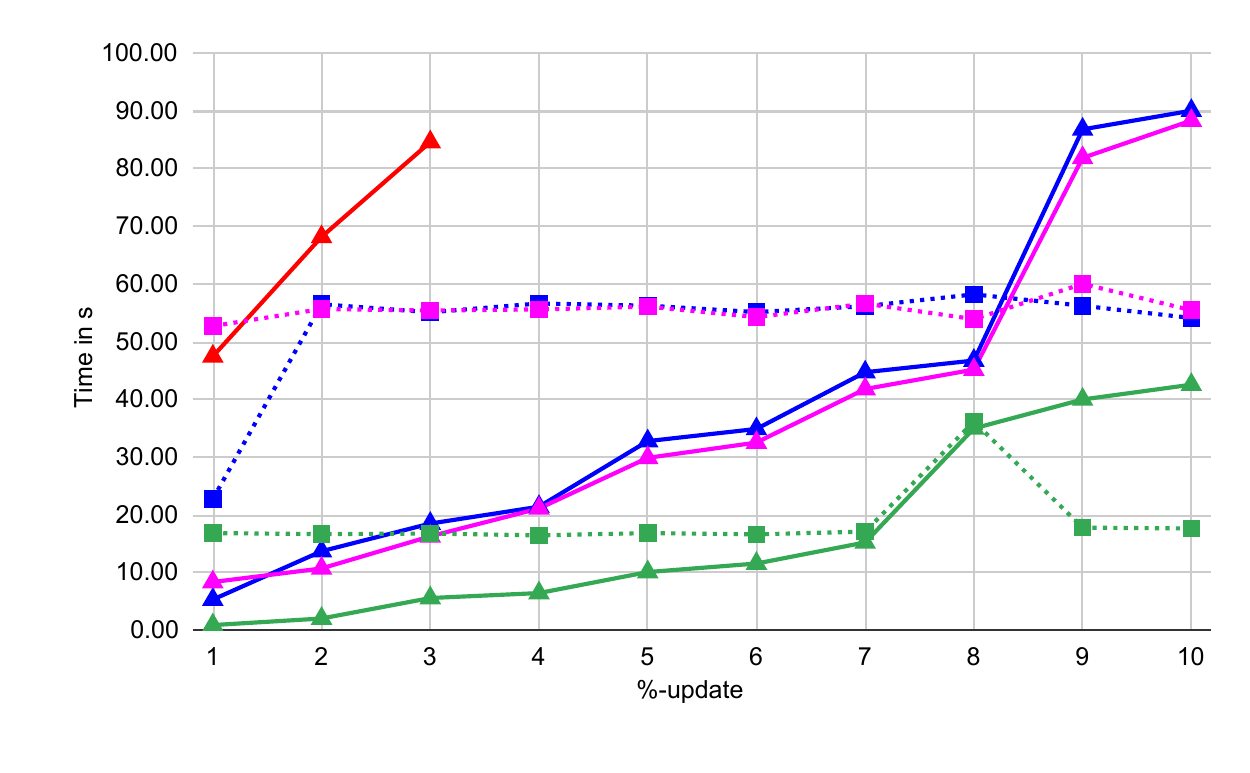}\\[-0.25ex]
{\small (d) Wikipedia Dataset}
\end{minipage}\hfill
\begin{minipage}{0.32\textwidth}\centering
\includegraphics[width=\linewidth,height=0.18\textheight,keepaspectratio]{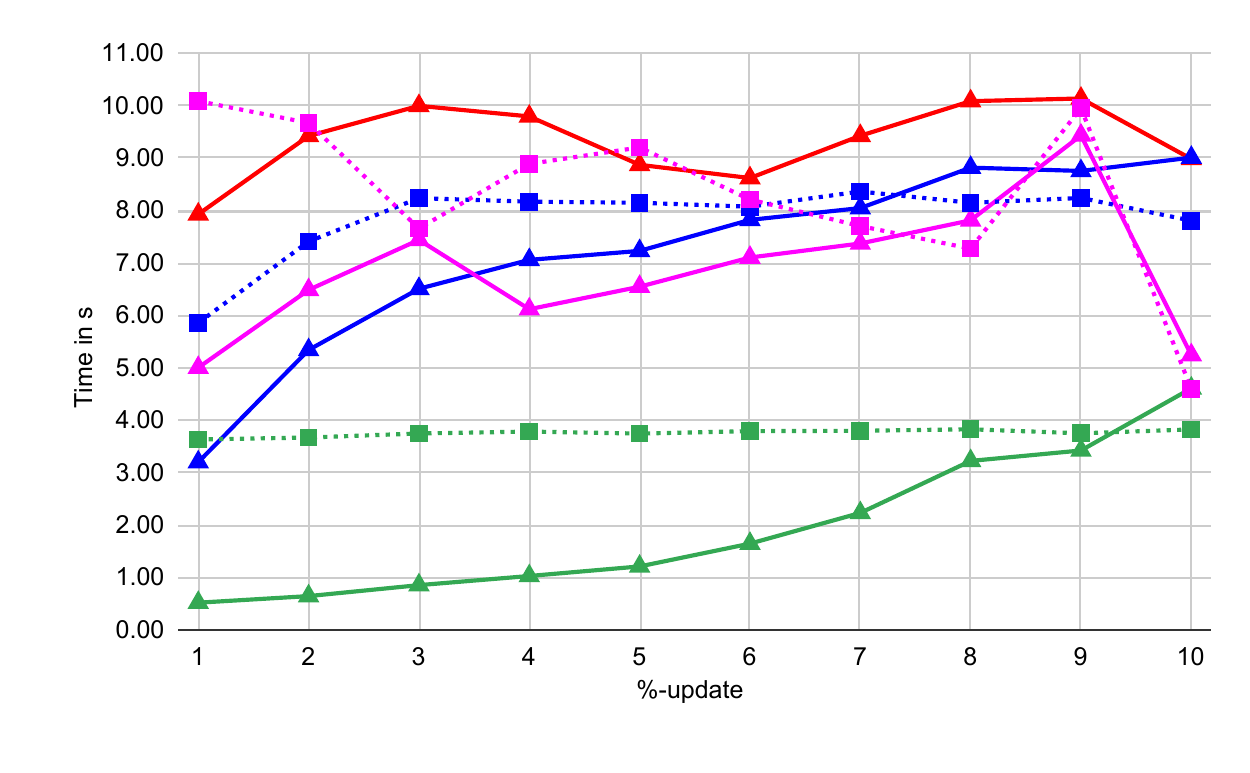}\\[-0.25ex]
{\small (e) Stack Overflow Dataset}
\end{minipage}\hfill
\begin{minipage}{0.15\textwidth}\centering
\includegraphics[width=\linewidth,keepaspectratio]{plots-paper/legend.png}\\[-0.25ex]
{\small Legend}
\end{minipage}

\vspace{0.5ex}
\captionof{figure}{Performance for mixed updates across all datasets.
\textit{Legend:} 
\textit{alt-pp} — state-of-the-art dynamic algorithm baseline; 
\textit{dyn-topo} — Dynamic \textit{Push–Relabel} algorithm with topology-based processing; 
\textit{dyn-data} — Dynamic \textit{Push–Relabel} algorithm with data-driven processing; 
\textit{dyn-pp-str} — Dynamic \textit{Push–Pull} algorithm with stream-based data processing; 
\textit{static-topo} — Static \textit{Push–Relabel} algorithm with topology-based processing; 
\textit{static-data} — Static \textit{Push–Relabel} algorithm with data-driven processing; 
\textit{static-pp} — Static \textit{Push–Pull} algorithm with data-driven processing.}
\label{fig:mixed}
\end{strip}
\FloatBarrier

\subsubsection{Decremental Updates}
The plots in Figure \ref{fig:dec} shows the performance of the dynamic algorithm when all the updates in the batch are smaller than the original capacity.
Unlike Incremental updates the dynamic implementations are doing better than the static and alternate approaches up to a certain percent of updates for each dataset. This percentage is different for each dataset.  This highlights that dynamic implementations are effective for decremental updates up to a threshold. This is probably because the decremental updates involve correcting the original flow and redirecting through a new path while incremental is effectively finding new paths from s to t.  
Across all the datasets, O2: push pull streams optimisation seems to be the winner. Similarly the topology-driven and data-driven implementations are giving similar performance.

\subsubsection{Mixed Updates}
The plots in Figure \ref{fig:mixed} shows the performance of the dynamic algorithm when half of the updates are smaller than the original capacity and the rest half is larger.
The behavior of dynamic algorithm for mixed updates is similar to the incremental and decremental updates, thus demonstrating that the dynamic algorithm is robust for a mixed update and not just purely incremental or decremental applications.

\subsubsection{Observations}
\begin{itemize}
\item Dynamic processing is efficient as it significantly outperforms the static approach.
    \item As the batch size increases, the time taken for dynamic processing generally increases.
    \item At most plots, we can see that the dynamic time is initially lower than the static time but as the batch size increases, it too increases and at a point the static time performs better. 
    \item Certain graphs exhibit unexpected peaks, deviating from the typical increasing trend.
    \item These anomalies may be attributed to:
    \begin{itemize}
        \item Irregularities in workload distribution,
        \item Structural characteristics of the graphs,
        \item Variations in the capacity updates,
        \item Or a combination of the above factors.
    \end{itemize}
\end{itemize}

\subsubsection{Results}
\begin{itemize}
\item Across the plots, the dynamic time has proven to be more efficient for smaller batch sizes, effectively minimizing the processing time due to its adaptive nature.
    \item It is thus efficient to use dynamic algorithm when the updates are small or up to the threshold in which the static algorithm outperfoms. This threshold depends on the graph structure and the implementation approach taken.   
\end{itemize}

\section{Conclusion}\label{sec conclusion}

We  contributed a complete GPU version of
the push relabel with the global relabel heuristic. We then contributed a pull-relabel algorithm which could be used for incremental and decremental updates. We also contribute the dynamic GPU maxflow algorithm to harness the benefit of global relabel.
Furthermore, our implementations demonstrated significant performance improvements
over static processing of dynamic graphs.
The experimental evaluation of the generated dynamic graph algorithms suggests
that using specialized dynamic algorithms might be advantageous for evolving graphs especially in maxflow which is a computation heavy. 
\subsection{Future Work}
\noindent\textbf{Multi-GPU and Distributed Scaling:} 
Extending the algorithm to multi-GPU or cluster settings would enable processing of trillion-edge graphs and facilitate integration with large-scale graph-analytics pipelines.

\vspace{0.5em}
\noindent\textbf{Algorithmic Generalization:} 
The dynamic Push--Pull paradigm can be generalized to other flow-based problems such as minimum-cut, multi-commodity flow, or dynamic bipartite matching, broadening the framework's applicability.

\vspace{0.5em}
\noindent\textbf{Integration into Graph Frameworks:} 
Incorporating the proposed kernels into mainstream GPU graph-processing libraries (e.g., \textit{Gunrock}, \textit{cuGraph}, \textit{GraphBLAST}) could allow community adoption and facilitate comprehensive comparative benchmarking.

\begin{acks}
We gratefully acknowledge the use of the computing resources at HPCE, IIT Madras. This work is supported by India's National Supercomputing Mission grant CS1920/1123/MEIT/008606.
\end{acks}

\bibliographystyle{ACM-Reference-Format}
\bibliography{sample}

\end{document}